\documentclass[10pt]{article}
\usepackage[dvips]{graphicx}
\usepackage{subfigure}
\usepackage{accents}
\begin{document}
\begin{center}
\textbf{{\large $Om$ Diagnostic for Dilaton Dark Energy}}
\vskip
0.35 in
\begin{minipage}{4.5 in}
\begin{center}
{\small Z. G. Huang$^{1,~\dag}$ and H. Q. Lu$^2$ \vskip 0.12 in
\textit{ $^1$School of
Science,~Huaihai~Institute~of~Technology,~Lianyungang,~China }}
\\
\textit{$^2$Department~of~Physics,~Shanghai~University,~Shanghai,~China
$^\dag$zghuang@hhit.edu.cn}
\end{center}
\vskip 0.2 in

{\small $Om$  diagnostic can differentiate between different models
of dark energy without the accurate current value of matter density.
We apply this geometric diagnostic to dilaton dark energy(DDE) model
and differentiate DDE model from LCDM. We also investigate the
influence of coupled parameter $\alpha$ on the evolutive behavior of
$Om$ with respect to redshift $z$. According to the numerical result
of $Om$, we get the current value of equation of state
$\omega_{\sigma0}$=-0.952 which fits the WMAP5+BAO+SN very well.
}
\end{minipage}
\end{center}
\vskip 0.2 in
\begin{flushleft}\textbf{1. Introduction}\end{flushleft}
\par So far, many astronomy observations including SNe Ia[1], SDSS[2], WMAP[3]
 provide us such a clear outline of the Universe: It is flat and full of an an unclumped form of energy density
pervading the Universe. The unclumped energy density called "Dark
Energy"(DE) with negative pressure, attributes to about 74 percent
of the total energy density. The remainder 26 percent of energy
density consists of matter including about 22 percent dark matter
density and about 4 percent baryon matter density. Beside this, we
know little about nature of DE. So, understanding the nature of Dark
Energy is one of most challengeable problem for modern astrophysics
and cosmology.
\par As the candidates of DE model, Quintessence[4], Phantom[5], Holographic Dark Energy[6], K-essence[7] and Quintom[8] so on,
have been being studied widely by many authors. Of course, the most
possible and fundamental candidate of DE is cosmological constant
with equation of state(EOS) $\omega=-1$. However, the cosmological
constant model suffers from two serious issues: Why the value of
cosmological constant $\Lambda$ is so tiny and not zero which is
called "fine-tuning problem". Why the energy density of $\Lambda$ is
just comparable with the matter energy density in recent time which
is called "coincidence problem". Alternative to the cosmological
constant include scalar field models called Quintessence which have
EOS $\omega>-1$, as well as more exotic "phantom" models with EOS
$\omega<-1$. The essential characteristics of these dark energy
models are contained in the parameter of its equation of state,
$p=\omega\rho$, where $p$ and $\rho$ denote the pressure and energy
density of dark energy, respectively, and $\omega$ is EOS parameter.
Quintessence model has been widely studied, and its EOS
$\omega_\phi$, is greater than $-1$. Such a model for a broad class
of potentials can give the energy density converging to its present
value for a wide set of initial conditions in the past and posses
tracker behavior. The quintessence potential $V(\phi)$ and the
equation of state $\omega_\phi(z)$ may be reconstructed from
supernova observations[9].
\par In our previous papers[10], we have successfully constructed dilaton dark energy(DDE) model
where we consider dilaton as a scalar field. For EOS
$\omega_\sigma>-1$ DDE model can be regarded as nonminimal coupled
Quintessence model while For EOS $\omega_\sigma<-1$ DDE model can be
regarded as nonminimal coupled Phantom model. Based on this, we
investigated the existence and stability of attractor solutions and
obtain that DDE model would admit a late time De sitter attractor
solution. Furthermore, parametrization of dark energy function, the
influence of dilaton scalar potential on the evolutive behavior of
attractor and reconstruction of scalar potential without dependence
on model were studied widely by us. In this paper, we will apply a
new geometric method--$Om$ diagnostic to DDE model.
\par $Om$, is constructed from the Hubble parameter $H=\frac{\dot{a}}{a}$
determined directly from observational data and provides a
\textit{null test} of the LCDM hypothesis. Here a(t) is the scale
factor of a Friedmann-Robertson-Walker(FRW) cosmology. In this paper
we will show that $Om$ is able to distinguish dynamical DDE from the
cosmological constant in a robust manner both with and without
reference to the value of the matter density, which can be a
significant source of uncertainty for cosmological reconstruction.
In other words, whether we know the current value of matter density
or not, we can distinguish DDE model from LCDM even other dark
energy models. The $Om$ diagnostic is in many respects the logical
companion to another geometric diagnostic--statefinder
$r\equiv\frac{\dddot{a}}{aH^3}$ where $r=1$ for LCDM while $r\neq1$
for evolving DE models. Hence $r(z_1)-r(z_2)$ provides a null test
for the cosmological constant. Similarly, the unevolving nature of
$Om(z)$ in LCDM furnishes $Om(z_1)-Om(z_2)$ as a null test for the
cosmological constant. Like the statefinder, $Om$ depends only upon
the expansion history of our Universe. However, while the
statefinder $r$ involves the third derivative of the expansion
factor a(t), $Om$ depends upon its first derivative only. Therefore,
$Om$ is much easier to reconstruct from observations.
\par This paper is organized as follows: Basic equations
of DDE model and introduction to $Om$ diagnostic are firstly
introduced in Sec.II. Based on these, we differentiate DDE model
from LCDM and investigate the influence of coupling parameter
$\alpha$ on the $Om_{DDE}$. These results are shown in figures
mathematically. Sec.III is conclusions.

\vskip 0.2 in
\begin{flushleft}\textbf{2. $Om$ Diagnostic For DDE Model}\end{flushleft}
\par Now let us consider the action of the Weyl-scaled induced gravitational theory:
\begin{equation}S=\int{d^4X\sqrt{-g}[\frac{1}{2}R(g_{\mu\nu})-\frac{1}{2}g^{\mu\nu}\partial_\mu\sigma\partial_\nu\sigma-W(\sigma)+L_{fluid}(\psi)}]\end{equation}
where
$L_{fluid}(\psi)=\frac{1}{2}g^{\mu\nu}e^{-\alpha\sigma}\partial_\mu\psi\partial_\nu\psi-e^{-2\alpha\sigma}V(\psi)$,
$\alpha=\sqrt{\frac{\kappa^2}{2\varpi+3}}$ with $\varpi$ being an
important parameter in Weyl-scaled induced gravitational theory,
$\sigma$ is dilaton field, $W(\sigma)$ is dilaton scalar potential,
$g_{\mu\nu}$ is the Pauli metric which can really represent the
massless spin-two graviton and should be considered to be physical
metric[11]. We work in units($\kappa^2\equiv8\pi G=1$). When
$W(\sigma)=0$, Weyl-scaled induced gravitational theory will reduce
to the Einstein-Brans-Dicke theory. We consider dilaton field as the
candidate of DE and call Weyl-scaled induced gravitational theory as
dilaton dark energy(DDE) model.
\par In Friedmann-Robertson-Walker universe, the field equations become:
\begin{equation}H^2=\frac{1}{3}[\rho_\sigma+e^{-\alpha\sigma}\rho_m]\end{equation}
\begin{equation}\frac{\ddot{a}}{a}=-\frac{1}{6}(e^{-\alpha\sigma}\rho_m+\rho_\sigma+p_\sigma)\end{equation}
\begin{equation}\ddot{\sigma}+3H\dot{\sigma}+\frac{dW(\sigma)}{d\sigma}=\frac{1}{2}\alpha e^{-\alpha\sigma}\rho_m\end{equation}
where $H=\frac{\dot{a}}{a}$ is Hubble parameter, $\rho_\sigma$ and $\rho_m=\rho_{m_0}
\frac{e^{\frac{1}{2}\alpha\sigma}}{a^3}$ are dark energy density and matter energy density respectively. The effective
energy density $\rho_\sigma$ and the effective pressure $p_\sigma$
of dilaton field can be expressed as follows
\begin{equation}\rho_{\sigma}=\frac{1}{2}\dot{\sigma}^2+W(\sigma)\end{equation}
\begin{equation}p_{\sigma}=\frac{1}{2}\dot{\sigma}^2-W(\sigma)\end{equation}
We can rewrite Eq.2 as follows:
\begin{equation}H^2=H^2_0[(1-\Omega_{m0})E(z)+\Omega_{m0}e^{-\frac{1}{2}\alpha\sigma}(1+z)^3]\end{equation}
where $\Omega_{m0}\equiv\rho_{m0}/3H^2_0$ is matter density parameter, and $E(z)$ is function of dark energy.
\par Now we introduce $Om$ geometric diagnostic[12] which has been
studied by many authors[13]
\begin{equation}Om(x)\equiv\frac{h^2(x)-1}{x^3-1},~~x=1+z,~~h^2(x)=\frac{H^2}{H_0^2}\end{equation}
For dark energy with a constant equation of state $\omega=const$,
\begin{equation}Om(x)=\Omega_{m0}+(1-\Omega_{m0})\frac{x^{3(1+\omega)}-1}{x^3-1}\end{equation}
We can easily find
\begin{equation}Om(x)=\Omega_{m0}\end{equation} in LCDM, whereas $Om(x)>\Omega_{m0}$ in quintessence ($\alpha>0$) while $Om(x)<\Omega_{m0}$ in quintessence ($\alpha<0$).
So, $Om(x)-\Omega_{m0}=0$ if candidate of DE is cosmological constant.
\par In this paper, we consider a simple form of dark energy function $E(z)=x^{3(1+\omega_\sigma)}$, so Hubble parameter $H$ can be expressed
\begin{equation}H^2(x)=H^2_0[(1-\Omega_{m0})x^{3(1+\omega_\sigma)}+\Omega_{m0}e^{-\frac{1}{2}\alpha\sigma}x^3]\end{equation}
where
$(1-\Omega_{m0})x^{3(1+\omega_\sigma)}+\Omega_{m0}e^{-\frac{1}{2}\alpha\sigma}x^3=h^2(x)$.
\par According to Eq.(8), we get the form of $Om$ in DDE model
\begin{equation}Om(x)_{DDE}=\frac{h^2(x)-1}{x^3-1}=\frac{(1-\Omega_{m0})x^{3(1+\omega_\sigma)}+\Omega_{m0}e^{-\frac{1}{2}\alpha\sigma}x^3-1}{x^3-1}\end{equation}
Comparing Eq.(9) and Eq.(12), we can see that coupling factor
$e^{-\frac{1}{2}\alpha\sigma}$ between dilaton field and matter can
affect the evolutive behavior of $Om_{DDE}$. When coupling parameter
$\alpha=0$, Eq.(12) reduces to Eq.(9). When $\alpha=0$ and
$\omega_\sigma=-1$, Eq.(12) reduces to Eq.(10). In other word,
$Om_{DDE}$ diagnostic can still provide a \textit{null test} of LCDM
when $\alpha=0$ and $\omega_\sigma=-1$ in DDE model.
\par We need deduce the expression of dilaton field $\sigma(z)$ if
we want to know the influence of coupling factor
$e^{-\frac{1}{2}\alpha\sigma}$ on the $Om$. So, according to Eq.5,
we have
\begin{equation}\frac{1}{2}\dot{\sigma}^2+[1+(\sigma-A)^2]e^{-B\sigma}=\rho_{\sigma0}z^{-3(1+\omega_\sigma)}\end{equation}
where we consider the dilaton scalar potential as the form
$W(\sigma)=[1+(\sigma-A)^2]e^{-B\sigma}$ with $A$ and $B$ being
constant. Many authors believe that field with this kind of
potential are predicted in the low energy limit of M-theory[14]. We
can rewrite Eq.(13) as follows
\begin{equation}\frac{1}{2}H^2(x)x^2(\frac{d\sigma}{dx})^2+[1+(\sigma-A)^2]e^{-B\sigma}=\rho_{\sigma0}x^{3(1+\omega_\sigma)}\end{equation}
\vskip 0.2 in
\begin{minipage}{0.4\textwidth}
\includegraphics[scale=0.8]{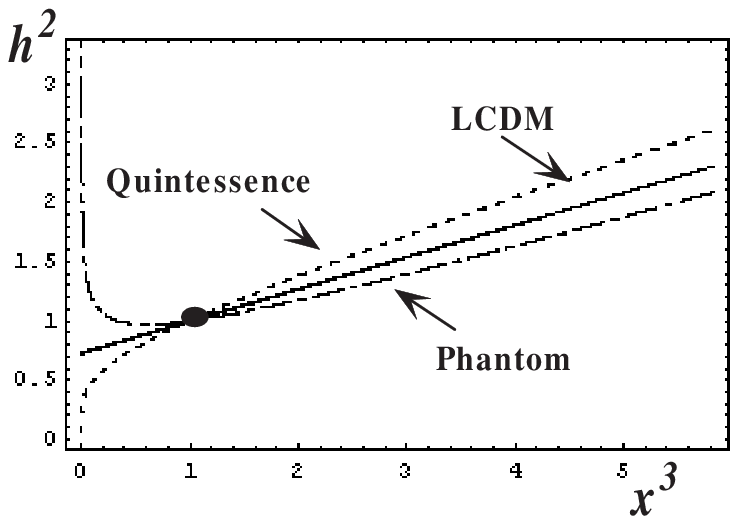}
{\footnotesize Fig.1 The evolutive trajectory of Hubble parameter square
$h^2(x)$ with respect to $x^3$(or $(1+z)^3$) for Quintessence(dot line), LCDM(real
line) and Phantom(dot-dashed line). We set $\Omega_{m0}=0.27$,
$\alpha=0.005$ and $\sigma_{z_0}=0.3$.}
\end{minipage}
\hspace{0.05\textwidth}
\begin{minipage}{0.4\textwidth}
\includegraphics[scale=0.58]{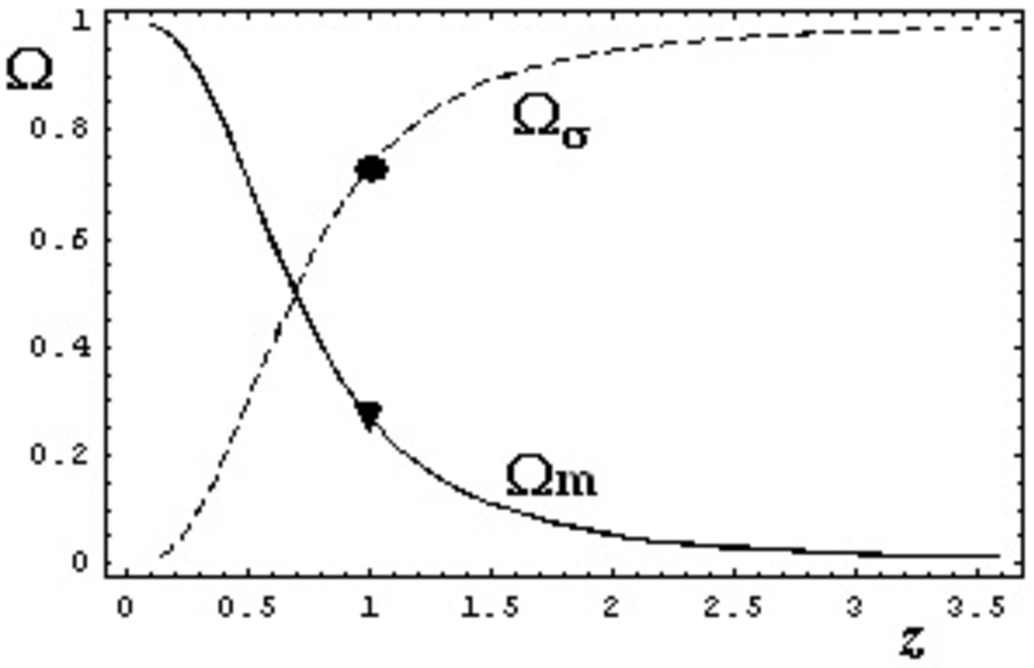}
{\footnotesize Fig.2 $\Omega_m$ and $\Omega_\sigma$ are plotted
against the red-shift $z$ for Quintessence. The filled Triangles and
the dot denote the current value of matter energy density parameter
$\Omega_{m0}$ and dark energy density parameter $\Omega_{\sigma0}$.}
\end{minipage}
\vskip 0.2 in
\par From Fig.1, we can see that the trajectory of $h^2(x)$
with respect to $x^3$ for LCDM is always a straight line in the
interval $-1<z<1.85$ whereas for Quintessence and Phantom the line
is curved in the interval $-1<z\ll1$. Clearly, a comparison of $Om$
at two different redshifts can lead to insights about the nature of
DE even if the value of $Om$ is not accurately known. Thus, the
two-point difference diagnostic
\begin{equation}Om(x_1,x_2)\equiv Om(x_1)-Om(x_2)\end{equation}
For LCDM
\begin{equation}Om(x_1)=Om(x_2)\end{equation}
So, $Om(x_1,x_2)=0$ if DE is a cosmological constant;
$Om(x_1,x_2)>0$ for quintessence while $Om(x_1,x_2)<0$ for phantom.
\par The evolutive trajectories of matter energy density parameter
$\Omega_m$ and $\Omega_\sigma$ are shown in Fig.2. The current value
of matter energy density parameter $\Omega_{m0}$ and dark energy
density parameter $\Omega_{\sigma0}$ are respective 0.27 and 0.73.
We can see that the DDE will evolve into de Sitter space-time at
late time. This result consists with the conclusion of our previous
paper[15].
\vskip 0.1 in
\begin{minipage}{0.4\textwidth}
\includegraphics[scale=0.8]{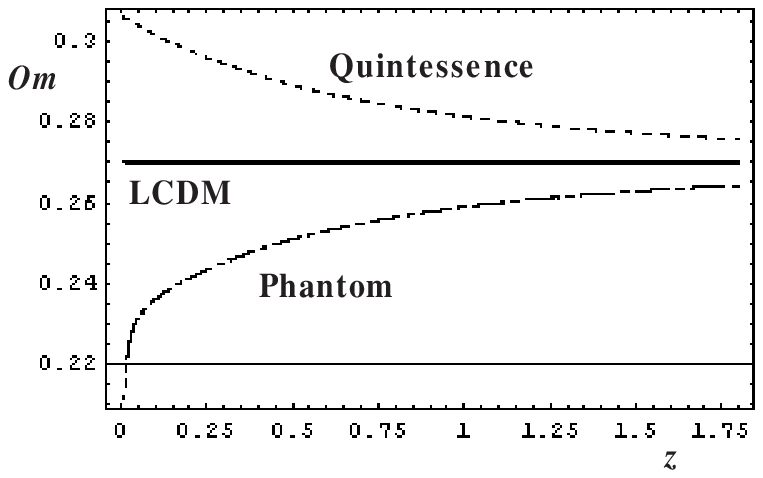}
{\footnotesize Fig.3 The evolutive trajectories of $Om$ with respect
to $z$ for Quintessence(dot line), Phantom(dot-dashed line) and
LCDM(real line). We set $\alpha=0.05$, $\Omega_{m0}=0.27$.}
\end{minipage}
\hspace{0.05\textwidth}
\begin{minipage}{0.4\textwidth}
\includegraphics[scale=0.8]{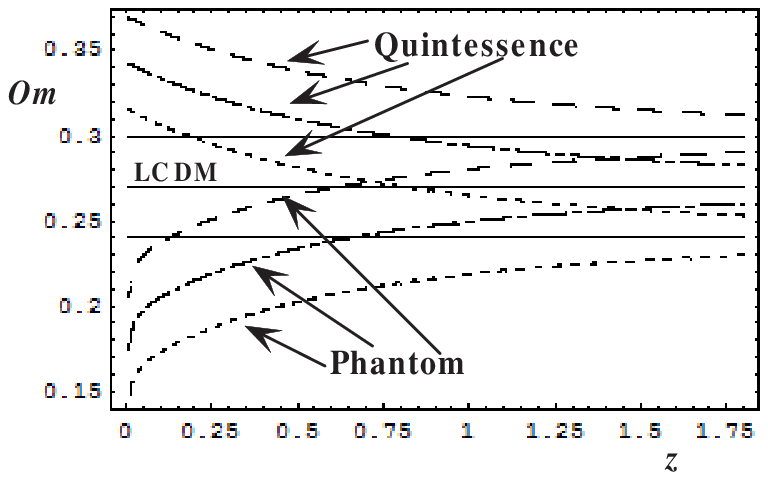}
{\footnotesize Fig.4 The evolutive trajectories of $Om$ with respect
to $z$ for Quintessence, Phantom and LCDM(real line) when we set
three different current values of $\Omega_{m0}$=0.27(dot-dashed
line), 0.24(dot line), 0.30(dashed line). We set $\alpha=0.05$.}
\end{minipage}
 \vskip 0.3 in
\begin{minipage}{0.4\textwidth}
\includegraphics[scale=0.8]{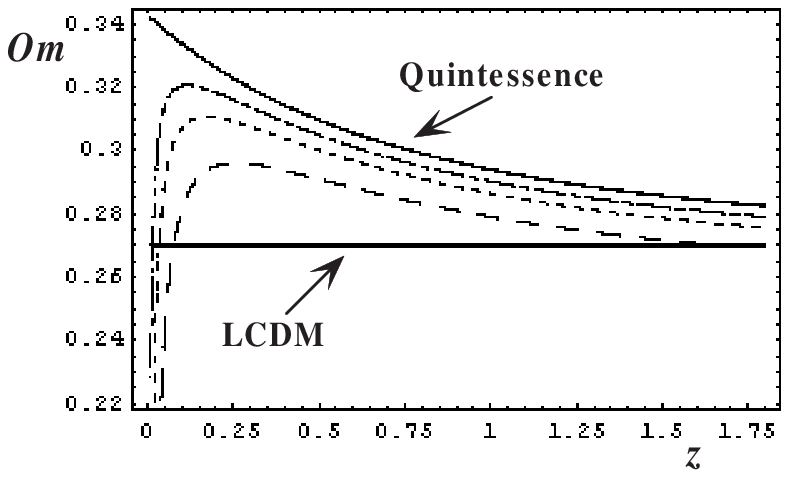}
\\
{\footnotesize Fig.5 The evolutive trajectories of $Om$ with respect
to $z$ for Quintessence when we set $\alpha$=0.000005(real line),
0.05(dot-dashed line), 0.1(dot line) and 0.2(dashed line). The
horizontal line corresponds to LCDM.}
\end{minipage}
\hspace{0.05\textwidth}
\begin{minipage}{0.4\textwidth}
\includegraphics[scale=0.8]{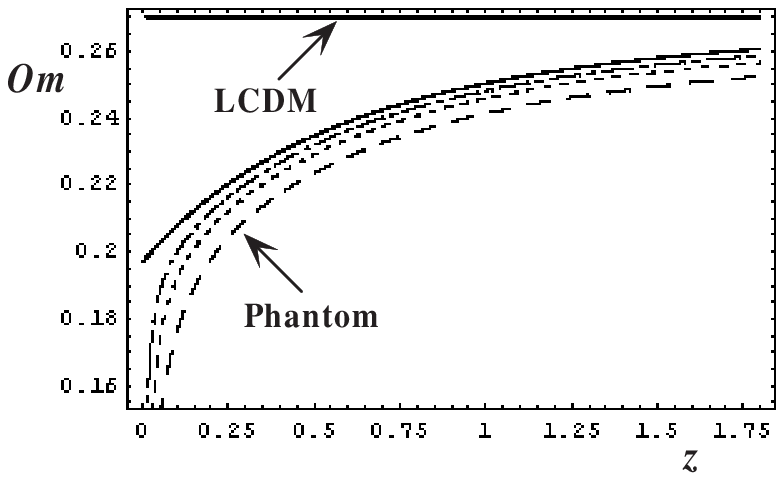}
\\
{\footnotesize Fig.6 The evolutive trajectories of $Om$ with respect
to $z$ for Phantom, when we set $\alpha$=0.000005(real line),
0.05(dot-dashed line), 0.1(dot line) and 0.2(dashed line). The
horizontal line corresponds to LCDM.}
\end{minipage}
\vskip 0.3 in
\par According to Fig.3, we can see clearly that the evolutive
behaviors of $Om$ with respect to $z$ for Quintessence, Phantom and
LCDM are very different. In the interval $0<z<1.85$, the slope of
$Om$ for Quintessence is negative while the slope of $Om$ for
Phantom is positive. The horizontal straight line corresponds to
LCDM. Therefore, we can easily distinguish Quintessence and Phantom
model from LCDM by the trajectories of their $Om$(slope of $Om$).
Furthermore, Fig.4 shows that the trajectories of $Om$ for
Quintessence, Phantom and LCDM when we set different values
$\Omega_{m0}$=0.27, 0.24 and 0.32. Clearly, whether we set a correct
current value of $\Omega_{m0}$=0.27 or incorrect values of
$\Omega_{m0}$=0.24 and 0.3, we can differentiate Quintessence and
Phantom from LCDM. The trajectories of $Om$ for Quintessence,
Phantom and LCDM$_{like}$ when $\Omega_{m0}$=0.24 and 0.3 can be
regarded as the results that Quintessence, Phantom and
LCDM($\Omega_{m0}=0.27$) moves up($\Omega_{m0}=0.3$) and
down($\Omega_{m0}=0.24$) paralleled. So, the slope of $Om$ can
differentiate between different models including Quintessence,
Phantom and LCDM, even if the value of the matter density is not
accurately known.
\par Fig.5 and Fig.6 show the influence of coupling parameter
$\alpha$ on the $Om$ diagnostic for Quintessence and Phantom
respectively. In Fig.5, we can see that the trajectory of $Om$ for
Quintessence moves upward with decrease of $\alpha$ from 0.2 to
0.00005. So, the asymptote of trajectory of $Om$ with
$\alpha\rightarrow0$ is the horizontal straight line which
corresponds to LCDM. Similarly, when the coupling parameter $\alpha$
changes from 0.2 to 0.00005, the trajectory of $Om$ for Phantom
moves upward too. When the the couple parameter
$\alpha\rightarrow0$, LCDM is the the asymptote of trajectory of
$Om$ for Phantom. \vskip 0.2 in
\begin{flushleft}\textbf{3. Conclusions}\end{flushleft}
\par In this paper we apply the $Om$ diagnostic to DDE model. We
have demonstrated that the plot of $h^2(x)$-$x^3$ for
Quintessence($\omega_\sigma>-1$), Phantom($\omega_\sigma<-1$) and
LCDM($\omega_\sigma=-1$). We can obtain that DDE model will evolve
into de Sitter space-time for Quintessence($\omega_\sigma>-1$) at
late time or "Big Rip" future singularity for
Phantom($\omega_\sigma<-1$) as $z\rightarrow0(x\rightarrow1)$. Fig.2
also shows DDE model in Quintessence admits a late time de Sitter
attractor solution. According to the expression
$\frac{Om(z)-\Omega_{m0}}{1-\Omega_{m0}}\simeq1+\omega_{\sigma0}+\frac{\Omega_{m0}(e^{-\frac{1}{2}\alpha\sigma_0}-1)}{1-\Omega_{m0}}$
when $x\rightarrow1$($z\rightarrow0$) and the numerical results of
differential equation Eq.(14), it is can be found that current EOS
of DDE $\omega_{\sigma0}\simeq-0.952$ which fits the combination
WMAP5+BAO+SN($\omega=-0.992\pm_{0.062}^{0.061}$) well.
\par We have also plotted the trajectory of $Om$ with respect to $z$
for Quintessence, Phantom and LCDM. We can easily distinguish DDE
model from LCDM according to the slope of their evolutive
trajectories. Horizontal straight line corresponds to LCDM while the
slope of $Om$ for Quintessence is negative and the slope of $Om$ for
Phantom is positive. The detailed numerical investigations show that
we can differentiate DDE models from LCDM even if the value of the
matter density is not accurately known.
\par At last, we investigate the influence of coupling parameter
$\alpha$ on $Om$ for DDE model. With the decrease of $\alpha$, the
shape of trajectory of $Om$ with respect to $z$ for both
Quintessence and Phantom, becomes more and more close to LCDM which
is the asymptote of trajectory of $Om$ with $\alpha\rightarrow0$.
The recent observational limit to coupling parameter $\alpha$ is
$\alpha<0.001$[16]. According to the numerical results, the
influence of coupling parameter $\alpha$ on $Om$ is very tiny in
observational range, which consists with our previous results.
\begin{flushleft}\textbf{Acknowledgements}\end{flushleft}
\par This work is partially supported by National Nature Science
Foundation of China under Grant No.10573012 and Natural Science
Foundation of Jiangsu Province under Grant No.07KJD140011.

\begin{flushleft}{\noindent\bf References}
 \small{
\item{1.}{ A. G. Riess et al., \textit{Astrophys. J}\textbf{607}, 665(2004);
\\ \hspace{0.15 in}A. G. Riess, \textit{Astron. J}\textbf{116}, 1009(1998);
\\ \hspace{0.15 in}S. Perlmutter et al., \textit{Astrophys. J}\textbf{517}, 565(1999);
\\ \hspace{0.15 in}N. A. Bahcall et al., \textit{Science}\textbf{284}, 1481(1999).}
\item{2.}{ M. Tegmark, et al., \textit{Phys. Rev. D}\textbf{69}, 103510(2004);}
\item{3.}{ C. L. Bennett et al., \textit{Astrophys. Phys. Lett}\textbf{148}, 1(2003);
\\\hspace{0.15 in}Hinshaw G et al., arXiv: 0803.0732;
\\\hspace{0.15 in}Nolta M R et al., arXiv: 0803.0593.}
\item{4.}{ J. S. Bagla, H. K. Jassal and T. Padmamabhan, \textit{Phys. Rev. D}\textbf{67}, 063504(2003);
\\\hspace{0.15 in}L. Amendola, M. Quartin, S. Tsujikawa and I. Waga, \textit{Phys.Rev.D}\textbf{74}, 023525(2006);
\\\hspace{0.15 in}E. Elizalde, S. Nojiri and S. D. Odintsov, arXiv:hep-th/0405034;
\\\hspace{0.15 in}S. Nojiri and S. D. Odintsov \textit{Phys. Lett. B}\textbf{639}, 144(2006);
\\\hspace{0.15 in}C. Wetterich \textit{Nucl. Phys. B}\textbf{302}, 668(1998);
\\\hspace{0.15 in}E. J. Copeland, M. Sami and S. Tsujikawa, arXiv:hep-th/060305;
\\\hspace{0.15 in}T. Padmanabhan, and T. R. Choudhury, \textit{Phys. Rev. D}\textbf{66}, 081301(2002);
\\\hspace{0.15 in}A. Sen, \textit{JHEP} \textbf{0204}, 048(2002);
\\\hspace{0.15 in}C. Armendariz-Picon, T. Damour and V. Mukhanov, \textit{Phys. Lett. B}\textbf{458}, 209(1999);
\\\hspace{0.15 in}A. Feinstein, \textit{Phys. Rev. D}\textbf{66}, 063511(2002);
\\\hspace{0.17 in}M. Fairbairn and M. H. Tytgat, \textit{Phys. Lett. B}\textbf{546} 1(2002);
\\\hspace{0.15 in}A. Frolov, L. Kofman and A. Starobinsky, \textit{Phys.Lett.B} \textbf{545}, 8(2002);
\\\hspace{0.17 in}L. Kofman and A. Linde, \textit{JHEP}\textbf{0207}, 004(2004);
\\\hspace{0.15 in}C. Acatrinei and C. Sochichiu, \textit{Mod. Phys. Lett. A}\textbf{18}, 31(2003);
\\\hspace{0.17 in}S. H. Alexander, \textit{Phys. Rev. D}\textbf{65}, 0203507(2002).}
\item{5.}{ H. Q. Lu, \textit{Int. J. Mod. Phys. D}\textbf{14}, 355(2005);
\\\hspace{0.17 in}W. Fang, H. Q. Lu, Z. G. Huang and K. F. Zhang, \textit{Int. J. Mod. Phys. D}\textbf{15}, 199(2006);
\\\hspace{0.17 in}T. Chiba, T. Okabe and M. Yamaguchi, \textit{Phys. Rev. D}\textbf{62}, 023511(2000).
\\\hspace{0.17 in}L. Amendola, S. Tsujikawa, and M. Sami, \textit{Phys. Lett. B}\textbf{632}, 155(2006);
\\\hspace{0.17 in}P. Singh, M. Sami and N. Dadhich, \textit{Phys.Rev. D}\textbf{68}, 023522(2003).}
\item{6.}{ M. Li, \textit{Phys. Lett. B}\textbf{603} 1(2004), arXiv:hep-th/0403127;
\\\hspace{0.17 in}Q. G. Huang and M. Li, \textit{JCAP}\textbf{0408}, 013(2004);
\\\hspace{0.17 in}M. Ito, \textit{Europhys. Lett}.\textbf{71}, 712-715(2005);
\\\hspace{0.17 in}K. Ke and M. Li, \textit{Phys.Lett.B}\textbf{606}, 173-176(2005);
\\\hspace{0.17 in}Q. G. Huang and M. Li, \textit{JCAP}\textbf{0503}, 001(2005);
\\\hspace{0.17 in}Y. G. Gong, B. Wang and Y. Z. Zhang, \textit{Phys. Rev. D}\textbf{72}, 043510(2005);
\\\hspace{0.17 in}X. Zhang, \textit{Int. J. Mod. Phys. D}\textbf{14}, 1597-1606(2005).}
\item{7.}{ C. Armend\'{a}riz-Pic\'{o}n, V. Mukhanov and P. J. Steinhardt, \textit{Phys.Rev.Lett}\textbf{85}, 4438(2000);
\\\hspace{0.17 in}C. Armend\'{a}riz-Pic\'{o}n, V. Mukhanov and P. J. Steinhardt, \textit{Phys. Rev. D}\textbf{63}, 103510(2001);
\\\hspace{0.17 in}T. Chiba, \textit{Phys.Rev.D}\textbf{66}, 063514(2002);
\\\hspace{0.17 in}M. Malquarti, E. J. Copeland, A. R. Liddle and M. Trodden, \textit{Phys. Rev. D}\textbf{67}, 123503(2003);
\\\hspace{0.17 in}R. J. Sherrer, \textit{Phys. Rev. Lett}\textbf{93)}, 011301(2004);
\\\hspace{0.17 in}L. P. Chimento, \textit{Phys. Rev. D}\textbf{69}, 123517(2004);
\\\hspace{0.17 in}A. Melchiorri, L. Mersini, C. J. Odman and M. Trodden, \textit{Phys. Rev. D}\textbf{68}, 043509(2003).}
\item{8.}{ W. Hao, R. G. Cai and D. F. Zeng, \textit{Class.Quant.Grav}\textbf{22}, 3189(2005);
\\\hspace{0.17 in}Z. K. Guo, Y. S. Piao, X. M. Zhang, Y.Z. Zhang, \textit{Phys.Lett. B}\textbf{608}, 177(2005);
\\\hspace{0.17 in}B. Feng, arXiv:astro-ph/0602156.}
\item{9.}{ M. Kowalski et al., arXiv:0804.4142(astro-ph).}
\item{10.}{Z. G. Huang and H. Q. Lu, \textit{Int. J. Mod. Phys. D}\textbf{15},1501(2006);
\\\hspace{0.17 in}Z. G. Huang, H. Q. Lu and W. Fang, \textit{Int. J. Mod. Phys. D}\textbf{16}, 1109(2007)[arXiv:hep-th/0610018];
\\\hspace{0.17 in}Z. G. Huang, X. H. Li and Q. Q. Sun, \textit{Astrophys. Space Sci.}\textbf{310},53(2007)[arXiv:hep-th/0610019].}
\item{11.}{Y. M. Cho, \textit{Phys. Rev.Lett}\textbf{68}, 3133(1992).}
\item{12.}{V. Sahni, A. Shafieloo and A. A. Starobinsky, \textit{Phys. Rev. D}\textbf{78}, 103502(2008)[arXiv:0807.3548(astro-ph)].}
\item{13.}{M. L. Tong and Y. Zhang, \textit{Phys. Rev. D}\textbf{80}, 023503(2009)[arXiv:0906.3646(gr-qc)];
\\\hspace{0.17 in}J. B Lu, L. X Xu, Y. X. Gui and B. R. Chang, arXiv:0812.2074(astro-ph).}
\item{14.}{C. Skordis and A. Albrecht, \textit{Phys. Rev. D}\textbf{66}, 043523(2002);
\\\hspace{0.17 in}A. Albrecht and C. Skordis, \textit{Phys. Rev. Lett}\textbf{84}, 2076(2000).}
\item{15.}{Z. G. Huang, H. Q. Lu and W. Fang, \textit{Class. Quantum. Grav}\textbf{23}, 6215(2006)[arXiv:hep-th/0604160].}
\item{16.}{T. Damour and K. Nordtvedt, \textit{Phys. Rev. Lett}\textbf{70}, 2217(1993).}
}
\end{flushleft}
\end{document}